\documentclass{raa}

\usepackage{graphicx,times}             
\input{epsf.sty}                        
\input{psfig.sty}                       

\def\Rs{R_{\odot}}

\begin{document}

   \title{A possible explanation of the Maunder minimum from a flux transport dynamo model}

   \volnopage{Vol.0 (200x) No.0, 000--000}      
   \setcounter{page}{1}           

   \author{Arnab Rai Choudhuri      
   \and Bidya Binay Karak
     }

   \institute{
            Department of Physics, Indian Institute of Science, Bangalore-560012; {\it
             arnab@physics.iisc.ernet.in}
      }

   \date{Received~~2008 month day; accepted~~2008~~month day}

   \abstract{
We propose that the poloidal field at the end of the last sunspot
cycle before the Maunder minimum fell to a very low value due to
fluctuations in the Babcock--Leighton process.  With this assumption,
a flux transport dynamo model is able to explain various aspects
of the historical records of the Maunder minimum remarkably well
on choosing the parameters of the model suitably to give the correct
growth time.
  }

   \authorrunning{A. R. Choudhuri \& B. B. Karak }            
   \titlerunning{Maunder minimum from flux transport dynamo }  

   \maketitle

%
%

\def\Rs{R_{\odot}}

\section{Introduction}


One of the most remarkable features of sunspot activity is the Maunder
minimum---a period during 1645--1715 when very few sunspots were seen.
Since sunspot activity has been unexpectedly low for more than a year
now, questions are raised whether we are on the threshold of another
such grand minimum.  The sunspot cycle is produced by the dynamo process
in the Sun.  It is important to understand the possible physical mechanism
which could have pushed the dynamo into the Maunder minimum.

Several authors have studied the archival records of sunspots during the
Maunder minimum (Sokoloff \& Nesme-Ribes 1994; Hoyt \& Schatten 1996).
The few sunspots seen during the Maunder minimum mostly appeared in
the southern hemisphere.  Sokoloff \& Nesme-Ribes (1994) have used the
archival data to construct a butterfly diagram for a part of the Maunder
minimum from 1670.  Usoskin, Mursula \& Kovaltsov (2000) argue that the
Maunder minimum started abruptly but ended in a gradual manner, indicating
that the strength of the dynamo must be building up as the Sun came out
of the Maunder minimum. When solar activity is stronger, magnetic fields
in the solar wind suppress the cosmic ray flux, reducing the production
of $^{10}$Be and $^{14}$C which can be used as proxies for solar activity.
From the analysis of $^{10}$Be abundance in a
polar ice core, Beer, Tobias \& Weiss (1998) concluded that the solar activity cycle 
continued during the Maunder minimum, although
the overall level of the activity was lower than usual. Miyahara et al.\
(2004) drew the same conclusion from their analysis of $^{14}$C abundance
in tree rings.

Beer, Tobias \& Weiss (1998) suggested that the nonlinearities in an
interface dynamo could be the cause of the Maunder minimum.  The flux
transport dynamo, in which the poloidal field is produced near the
solar surface by the Babcock--Lieghton mechanism and transport by
meridional circulation plays an important role, has emerged as an
attractive model for the solar cycle ever since the first two-dimensional
models were constructed by Choudhuri, Sch\"ussler \& Dikpati (1995)
and Durney (1995). Our aim is to explore whether the Maunder minimum
can be explained within the framework of the flux transport dynamo
model developed by our group (Nandy \& Choudhuri 2002; Chatterjee,
Nandy \& Choudhuri 2004).  

It is not yet established whether the irregularities in the solar cycle
are caused primarily by the nonlinearities (Beer, Tobias \& Weiss 1998)
or by stochastic fluctuations (Choudhuri 1992; Brandenburg \& 
Spiegel 2008). Charbonneau, Blais-Laurier \& St-Jean (2004) presented
flux transport dynamo simulations with stochastic fluctuations, which
led to intermittencies resembling the Maunder minimum. These simulations
used a rather low turbulent diffusivity of $1.67 \times 10^{11}$ cm$^2$
s$^{-1}$.  Such a low diffusivity makes the 
diffusive decay time or the `memory' of the dynamo very
long (of the order of a century) and most probably this long memory
played a role in producing intermittencies of similar duration. As
argued by Jiang, Chatterjee \& Choudhuri (2007), there are several
compelling reasons to believe that the diffusivity is at least one order
of magnitude larger, making the memory not longer than a few years. We
explore whether such a short-memory flux transport dynamo can give rise
to the Maunder minimum.

After discussing the methodology in \S2, we present the results in
\S3.  Our main conclusions are summarized in \S4.

\section{Methodology}
 
Choudhuri, Chatterjee
\& Jiang (2007) argue that the Babcock--Leighton mechanism, in which
the poloidal field is produced from the decay of tilted bipolar sunspots,
involves randomness because the convective buffeting on rising flux
tubes causes a scatter in the tilt angles (Longcope \& Choudhuri 2002).
The simplest way of treating this randomness is to introduce a fluctuation
in the poloidal field produced at the end of a cycle.  Although the statistics
may not yet be completely convincing, there is evidence that the polar
field at the end of a cycle has a correlation with the strength of the
next cycle.  Jiang, Chatterjee \& Choudhuri (2007, see also Yeates, Nandy
\& Mackay 2008) showed that a flux transport dynamo
model with a reasonably high turbulent diffusivity can explain this
correlation.  This observed correlation suggests that the polar field at
the end of the last cycle just before the Maunder minimum might have been
very low. We present a calculation based on this assumption.

In spite of the fluctuations in the Babcock--Leighton process, one may
wonder if there is a statistically significant probability of the
polar field falling to a very low value at the end of a cycle.  We
are right now carrying on a detailed analysis of that.  It is known
that a large bipolar sunspot pair can produce a very large perturbation
in the Babcock--Leighton process (Wang, Nash \& Sheeley 1989).  During
the decay phase of a cycle, if a very large bipolar sunspot pair 
emerges with a `wrong' tilt (i.e. with the following spot closer to
the equator), it may be possible for it to neutralize the polar field
which had been growing due to contributions from the earlier spots.
Even if such a thing happens in one hemisphere, it would seem extremely
improbable for this to happen in the two hemispheres simultaneously.
In the flux transport dynamo with a reasonably high diffusivity, however,
the two hemispheres remain coupled and the growth of large hemispheric
asymmetries are suppressed due to this diffusive coupling (Chatterjee
\& Choudhuri 2006; Goel \& Choudhuri 2009).  Observational data
also suggest that hemispheric asymmetries have never been very large
in the cycles during the last century (Goel \& Choudhuri 2009).
So, if the polar field becomes close to zero in one hemisphere,
it is possible that diffusive coupling will make it small in the
other hemisphere also. Representing the dynamo by a simple iterative
map, Charbonneau (2005) also concluded that a strong cross-hemispheric
coupling was essential for the production of the Maunder minimum.

We present calculations on the basis of the ansatz that the polar
field at end of the last cycle before the Maunder minimum fell to
rather low value in the northern hemisphere (like 0.1 of the typical
average value of the polar field at the end of a cycle)
and also became quite low in the
southern hemisphere due to the diffusive coupling between the
hemispheres.  We do not try to justify this ansatz further
in this paper.  A future paper will analyze this ansatz further and
will try to estimate a statistical probability of such a thing 
happening.  Here we merely show that this ansatz allows the flux
transport dynamo model to reproduce many features of the Maunder
minimum remarkably well.  

\section{Results}

We carry out our calculations with the solar dynamo code {\em Surya}, which 
is made available upon request. The details of our dynamo model can be found in 
Chatterjee, Nandy \&
Choudhuri (2004).  
It was suggested by Choudhuri, Chatterjee \& Jiang (2007) that the
cumulative effect of fluctuations in the Babcock--Leighton process
during the decaying phase of a cycle can be modelled by stopping the 
dynamo code at the end of a sunspot cycle and then multiplying the
poloidal field function $A (r, \theta)$ above $r = 0.8 \Rs$ by a factor $\gamma$.
Goel \& Choudhuri (2009) used two values of $\gamma$ for the two
hemispheres in order to model the hemispheric asymmetry. Following this
approach, we stop the code at a sunspot minimum after obtaining a
relaxed periodic solution.  Then, in the northern hemisphere, we multiply
$A(r, \theta)$ above $r = 0.8 \Rs$ by a factor $\gamma_N$, whereas
we use the factor $\gamma_S$ for the southern hemisphere. To get
good results, we have to make another assumption which we agree is
somewhat ad hoc.  One of the unsatisfactory aspects of our model so far
is that we have been unable to avoid some degree of overlap
between cycles (i.e.\ the previous cycle continues at low latitudes even
after the new cycle has started at high latitudes).  See Fig.~13 of Chatterjee,
Nandy \& Choudhuri (2004) or Fig.~3 of Choudhuri, Chatterjee \& Jiang (2007).
Although one sees some amount of overlap between cycles in observational
data also, our theoretical model has a much bigger overlap.
Since we do not want further eruptions to take place and produce more
poloidal field after we have set the poloidal field to low values, we
multiply the toroidal field $B (r, \theta)$
everywhere by a factor $\gamma_t = 0.8$ at the same time
when we change the poloidal field. As a sunspot  
eruption in our model takes place only when $B$ is above a
critical value $B_c$, this reduction in $B$ ensures that an
eruption will not take place for some time.

\begin{figure}
\centering{\includegraphics[width=15cm]{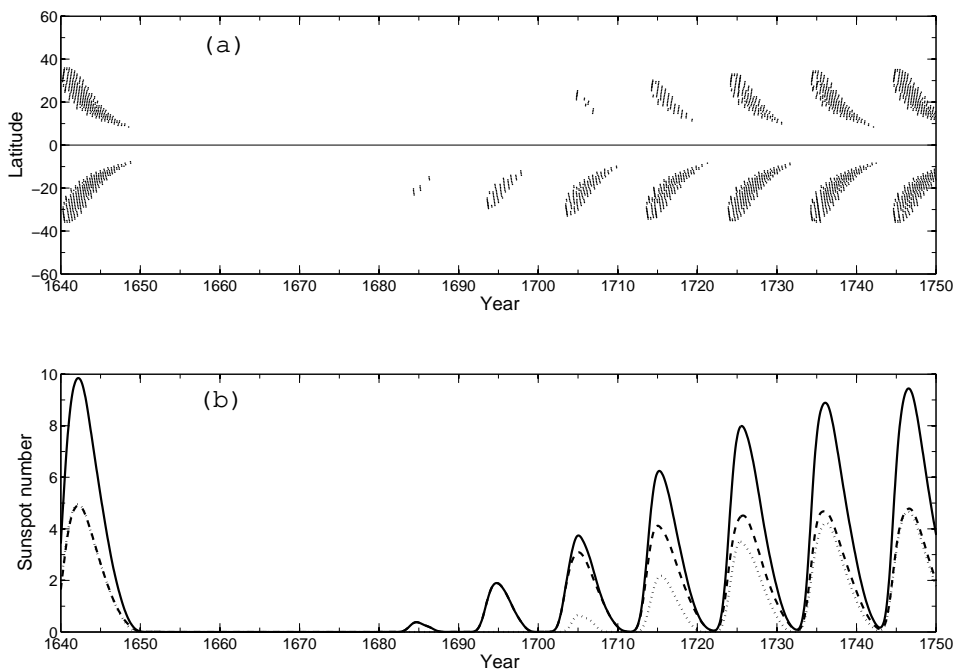}}

\caption{Theoretical results from our dynamo simulation covering
the Maunder minimum. (a) The theoretical butterfly diagram. (b) The
theoretical sunspot number (this is obtained by first considering
the monthly number of eruptions and then smoothing it by 
taking a running average over 5 months iteratively). The dashed and dotted lines show the
sunspot numbers in northern and southern hemispheres, whereas the solid
line is the total sunspot number.}
\end{figure}

After stopping the code at a sunspot minimum and making these changes,
we run the code for several cycles without any further interruption.
When we choose the same parameters as Choudhuri, Chatterjee \& Jiang (2007),
we find that the dynamo, which had the poloidal field reduced to very
low values at a minimum, bounces back to its original strength in
a growth time of about 35 yr.  In the case of a simple $\alpha \Omega$
dynamo, it can be analytically shown that the growth time is prolonged
on decreasing $\alpha$ or increasing the diffusivity $\eta$ (see, for
example, Choudhuri 1998, \S16.6). The same seems to hold for the more complex
flux transport dynamo as well.  We find that we get an appropriate growth
time if we reduce the amplitude of $\alpha$ from 25 m s$^{-1}$ to
21 m s$^{-1}$ and increase the turbulent diffusivity $\eta_p$ of the poloidal
field within the convection zone from $2.4 \times 10^{12}$ cm$^2$ s$^{-1}$
to $3.2 \times 10^{12}$ cm$^2$ s$^{-1}$. With these values of $\alpha$
and $\eta_p$, we use the poloidal field reduction factors $\gamma_N = 0.0$
and $\gamma_S = 0.4$ at a minimum.
Fig.~1(a) shows the theoretical butterfly diagram, whereas Fig.~1(b)
shows the sunspot number as a function of time.  In order to facilitate
comparison with observational data, we have taken the beginning of Fig.~1
to be the year 1640.  Hopefully all readers will agree that these
two figures are remarkably similar to the observational butterfly
diagram given in Fig.~1(a) of Sokoloff \& Nesme-Ribes (1994) and
the observational sunspot number plot given in Fig.~1 of Usoskin, Mursula \&
Kovaltsov (2000). 

Even though the diffusive decay time or the memory of our dynamo is
only a few years, it is possible to get a grand minimum of much longer duration
by making the growth time suitably large. The indication in the observational
data that solar activity at the beginning of the Maunder minimum decreased
abruptly and then built up gradually (Usoskin, Mursula \& Kovaltsov 2000)
strongly supports our interpretation that a grand minimum is caused
by a sudden decrease in the poloidal field due to fluctuations in the
Babcock--Leighton process and then the dynamo regains its strength in
its growth time.  If the duration of the grand minimum is really an
indication of the dynamo growth time, then it is possible to put constraints
on the parameters of the dynamo from this.  The combination of parameters
we used to produce Fig.~1 is by no means the only combination that gives
the correct duration of the grand minimum.  Table~1 lists some other combinations
which also produce grand minima resembling the Maunder minimum.  It may
be noted that $\alpha / \eta_p^2$ for all these combinations are comparable,
suggesting that the dynamo numbers (see, for example, Choudhuri 1998,
\S16.6) for these combinations are comparable.

\begin{table}
 \begin{center}
\begin{tabular}{ | l | c | c | c | c | l | }

\hline\hline
$\eta_p{\rm (cm^2/s)}$ & $\alpha({\rm m/s})$ & $\alpha/\eta_p^2$ & $\gamma{_t}$ & $\gamma{_N}$ & $\gamma{_S}$
 \\[1.5ex]
 \hline\hline

  $3.2\times10^{12}$ & 21 & 2.05 $\times10^{-24}$& 0.8 & 0.0 & 0.4 \\ [1.5ex]
 \hline

                                      &                      &                                        & 0.8 & 0.2 & 0.6 \\
 \raisebox{1.5ex}{$3.2\times10^{12}$} &\raisebox{1.5ex}{20}  &\raisebox{1.5ex}{1.95$\times10^{-24}$}  & 0.9 & 0.0 & 0.4 \\

 \hline
                                      &                      &                                        & 0.8 & 0.2 & 0.6 \\
 \raisebox{1.5ex}{$3.3\times10^{12}$} &\raisebox{1.5ex}{22}  &\raisebox{1.5ex}{2.02$\times10^{-24}$}  & 0.9 & 0.0 & 0.5 \\
 \hline

                                      &                      &                                        & 0.8 & 0.2 & 0.6 \\
 \raisebox{1.5ex}{$2.9\times10^{12}$} &\raisebox{1.5ex}{17}  &\raisebox{1.5ex}{2.02$\times10^{-24}$}  & 0.9 & 0.0 & 0.4 \\
 \hline
                                      &                      &                                        & 0.8 & 0.2 & 0.6 \\
 \raisebox{1.5ex}{$2.7\times10^{12}$} &\raisebox{1.5ex}{15}  &\raisebox{1.5ex}{2.06$\times10^{-24}$}  & 0.9 & 0.0 & 0.5 \\
 \hline

 \hline
                                      &                      &                                        & 0.8 & 0.2 & 0.6 \\
 \raisebox{1.5ex}{$2.2\times10^{12}$} &\raisebox{1.5ex}{11}  &\raisebox{1.5ex}{2.27$\times10^{-24}$}  & 0.9 & 0.0 & 0.4 \\
 \hline
  \end{tabular}
 \end{center}
\caption{A few combinations of parameters which produce figures similar to Figure~1}
\end{table}

Sunspots do not erupt in our model until the toroidal field builds up
to strength larger than $B_c$.
Even when there are few or no sunspots, the dynamo must be growing in
an oscillatory fashion.  In the absence of sunspots, the Babcock--Leighton
process is not possible.  As argued by Choudhuri (2003), the
toroidal field is expected to be concentrated in limited regions outside
which it may be diffuse.  Since flux tube simulations (Choudhuri 1989;
D'Silva \& Choudhuri 1993; Fan, Fisher \& DeLuca 1993) suggest that the
initial field strength inside flux tubes has to be of order $10^5$ G,
an eruption presumably takes place only when the concentrated field reaches
a value of $10^5$ G (corresponding to $B_c$ in our model).  If the toroidal
field does not become strong enough to produce sunspot eruptions, then
presumably this toroidal field is carried upward in the low-latitude regions
of the convection zone by the upflowing meridional circulation (and probably
also the uprising regions of convection).  The $\alpha$ coefficient in our
equations acting on this toroidal field can be interpreted as the traditional
$\alpha$ in mean field MHD (see Choudhuri 1998, \S16.5). 
Even when sunspot eruptions take place, Choudhuri
\& Dikpati (1999) concluded that the distribution of large-scale solar magnetic
field can be modelled particularly well by assuming the poloidal field to have
two sources---from Babcock--Leighton mechanism and from $\alpha$ effect
working on weaker subsurface toroidal field. When sunspots are absent,
the strength of the dynamo builds up due to the $\alpha$ effect operating
on the weaker toroidal field.

\begin{figure}
\centering{\includegraphics[width=12cm]{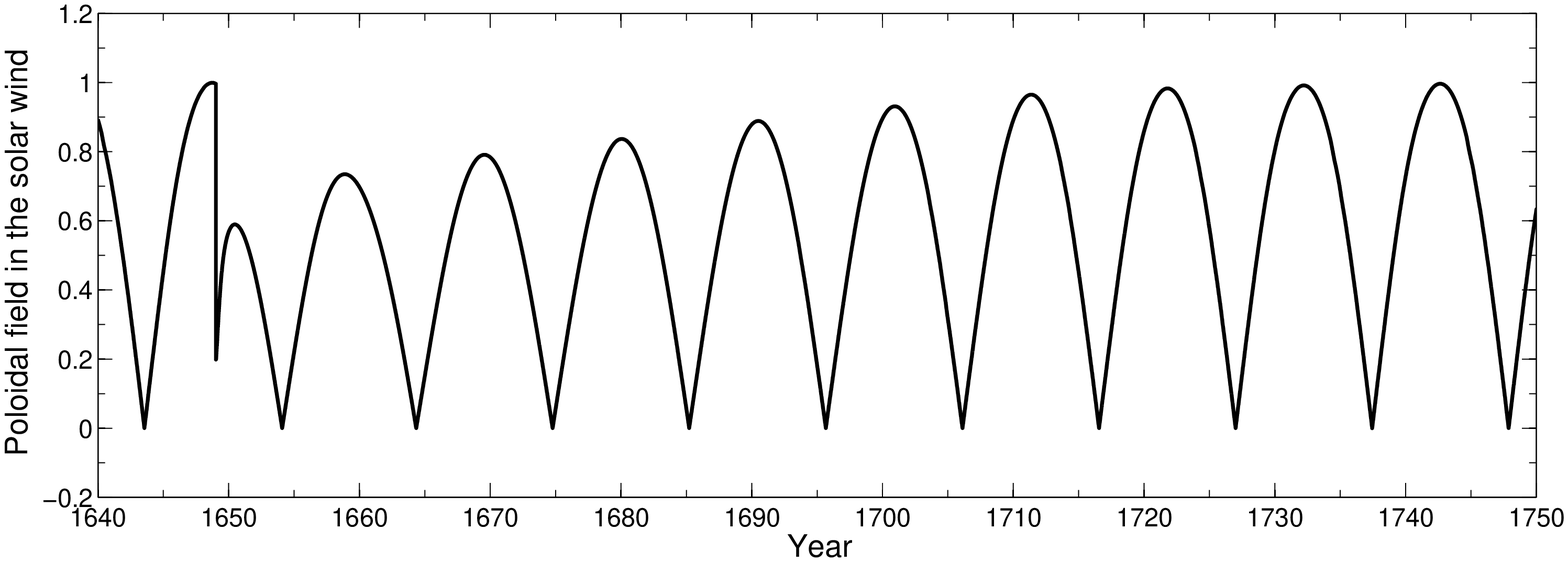}}

\caption{The theoretically calculated average radial magnetic field in the
solar wind (in arbitrary units) as a function of time.}
\end{figure}

The $^{10}$Be and $^{14}$C abundances depend on the magnetic field in the solar wind.
We have implemented the upper boundary condition in our model in such
a way that the magnetic field becomes radial beyond a distance of $3 R_{\odot}$
due to the stretching by the solar wind (see Dikpati \& Choudhuri 1995).  
At a distance where the magnetic
field is radial, its average strength can be obtained by averaging over
the spherical surface there, i.e.
$$<|B_r (r)|> = \frac{1}{2} \int_0^{\pi} |B_r (r, \theta)| \sin \theta \, d \theta.$$
Fig.~2 shows this as a function of time, the sudden change around 1648 being due to 
the change in the poloidal field.  Although the strength of $<|B_r (r)|> $ becomes 
weaker during the Maunder minimum, the oscillations in the 
magnetic field continue, explaining
why oscillations are seen in the $^{10}$Be and $^{14}$C abundance data during the Maunder
minimum. It is interesting to note that the magnetic field in the solar wind
in our model does not become that much weaker even when there are no sunspots.

\section{Conclusion}

We have shown that different characteristics of the Maunder minimum can be
explained very elegantly on the basis of the simple ansatz that, as a result of
randomness in the Babcock--Leighton process, the poloidal field 
at the end of the last cycle before the Maunder minimum fell to a rather
low value in the northern hemisphere and also became small in the southern
hemisphere due to the diffusive coupling between the hemispheres.  
While we are carrying on an analysis to show that this ansatz
can be justified on statistical grounds, the fact that calculations based
on this ansatz agree with observational data so extremely well gives credence
to this ansatz. Our model explains why the Maunder minimum started abruptly,
but ended with a more gradual growth of cycle strengths.  In the case of
the `failed' Dalton minimum (cycles 5 and 6) also, sunspot number data indicate
that the cycle strength fell abruptly and then grew more gradually.  The growth time
in this case was shorter than what it was in the case of Maunder minimum.
While we are identifying the duration of the Maunder minimum with the dynamo
growth time, it should be kept in mind that this is not a linear problem.
The toroidal field is not allowed to grow much beyond $B_c$ due to magnetic
buoyancy.  If the fall in dynamo strength is less abrupt, then the dynamo
bounces back in shorter time. Even to produce the Maunder minimum in our
model, we did not have to use unreasonably low polar field diminution factors
at the end of a cycle ($\gamma_N = 0.0$ and $\gamma_S = 0.4$ being our
choices).

We end with a comment whether we could be entering another grand minimum
right now.   Based on the polar field data provided by Svalgaard, Cliver
\& Kamide (2005), Choudhuri, Chatterjee \& Jiang (2007) estimated that 
the polar field diminution factor $\gamma$ at the
end of cycle~23 was 0.6.  According to our dynamo model, this drop
in the polar field is not sufficient to cause a grand minimum without
sunspot eruptions for many years. However, in view of the various uncertainties
in our model, we should keep our mind open and wait to see what the Sun
has in store for us.

\bigskip
{\it Acknowledgement.}
ARC acknowledges partial support from a DST project No.SR/S2/HEP--15/2007.

\end{document}